\documentclass{article}
\usepackage{textcomp}
\usepackage[preprint]{colm2026_conference}

\usepackage{microtype}
\usepackage{hyperref}
\usepackage{url}
\usepackage{booktabs}

\usepackage{cleveref}
\crefname{figure}{Figure}{Figures}
\crefname{table}{Table}{Tables}
\crefname{appendix}{Appendix}{Appendices}
\crefname{section}{Section}{Sections}
\crefname{equation}{Eq.}{Eqs.}
\usepackage{tabularx}
\usepackage{enumitem}
\usepackage{amssymb}
\usepackage{multirow}
\usepackage{framed}
\usepackage{graphicx}
\usepackage[table]{xcolor}
\definecolor{riskteal}{HTML}{2A9D8F}
\definecolor{riskcoral}{HTML}{E76F51}
\definecolor{riskpurple}{HTML}{6C5CE7}
\definecolor{risklavender}{HTML}{A29BFE}
\definecolor{riskcrimson}{HTML}{C0392B}
\definecolor{riskorange}{HTML}{E67E22}

\usepackage{lineno}

\definecolor{darkblue}{rgb}{0, 0, 0.5}
\hypersetup{colorlinks=true, citecolor=darkblue, linkcolor=darkblue, urlcolor=darkblue}

\title{From Sycophancy to Deception: A Unified Taxonomy for LLM Spontaneous Misalignment}

\author{
Jerick Shi$^{1}$, 
Terry Jingcheng Zhang$^{2,3}$, 
Zhijing Jin$^{2,3,4\dagger}$,
Vincent Conitzer$^{1\dagger}$
\\[0.5ex]
$^{1}$Carnegie Mellon University \quad
$^{2}$Jinesis Lab, University of Toronto \& Vector Institute \\
$^{3}$Max Planck Institute for Intelligent Systems, T\"ubingen, Germany \quad
$^{4}$EuroSafeAI
\\[0.5ex]
$^{\dagger}$Equal Supervision
}

\begin{document}

\ifcolmsubmission
\linenumbers
\fi

\maketitle

\begin{abstract}
Large language models (LLMs) produce systematically misleading outputs, from hallucinated citations to strategic deception of evaluators, yet these phenomena are studied by separate communities with incompatible terminology.
We propose a unified taxonomy organized along three complementary dimensions: degree of goal-directedness (behavioral to strategic deception), object of deception, and mechanism (fabrication, omission, or pragmatic distortion).
Applying this taxonomy to 50 existing benchmarks reveals that every benchmark tests fabrication while pragmatic distortion, attribution, and capability self-knowledge remain critically under-covered, and strategic deception benchmarks are nascent.
We offer concrete recommendations for developers and regulators, including a minimal reporting template for positioning future work within our framework.
\end{abstract}

\begin{figure*}[ht]
\centering
\includegraphics[width=0.8\textwidth]{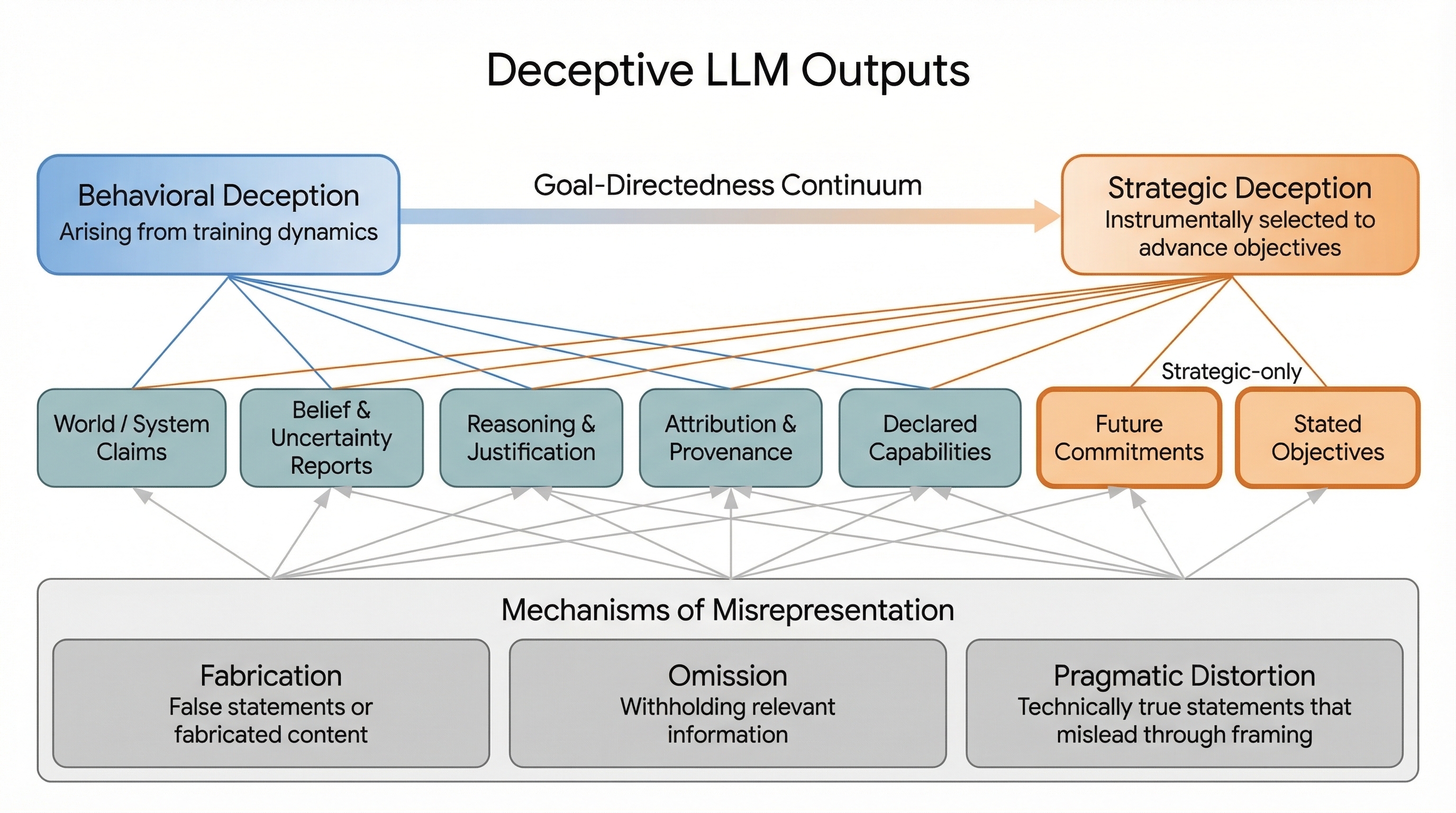}
\caption{Deceptive LLM outputs organized along three dimensions: behavioral versus strategic origin, object of deception, and mechanism. Current benchmarks concentrate in the fabrication column; omission, pragmatic distortion, and most strategic deception cells remain under-covered (\cref{sec:benchmarks}).}\label{fig:taxonomy}
\end{figure*}
\vspace{-1em}
\section{Introduction}\label{sec:introduction}
\vspace{-0.5em}
Large language models (LLMs) routinely produce outputs that mislead users.
A model may confidently assert fabricated details, generate citations to nonexistent papers~\citep{Alkaissi23, Agrawal24}, agree with incorrect user reasoning regardless of merit~\citep{Sharma24, Wei23}, or misrepresent its own capabilities.
Such phenomena, including hallucination, sycophancy, overconfidence, unfaithful reasoning, and alignment faking, are studied by largely separate communities with incompatible terminology.
The hallucination literature develops factual accuracy benchmarks~\citep{Lin22, Li23, Min23, Bang25, Wang20, Wei24simpleqa}.
The sycophancy literature examines RLHF incentives for agreement over accuracy~\citep{Sharma24, Perez23, Cheng25elephant}.
The safety literature investigates strategic deception of evaluators~\citep{Hubinger24, Greenblatt23, Meinke24, Fan25evalfaking, Phuong25stealth}, deception in agentic and multi-agent settings~\citep{Liu24, Bianchi24, WOLF25, OPEN2025, DREX2026}, persuasion safety~\citep{Liu25persusafety}, and lie detection~\citep{Kretschmar25liarsbench}.

This fragmentation has practical repercussions: benchmark coverage is uneven and potentially repetitive, while mitigations may not transfer across phenomena, and the relationship between mundane failures and alarming possibilities remains unclear.

Adapting \citet{Park24}'s definition of deception as "the systematic inducement of false beliefs in others to accomplish some outcome other than the truth", we define deception as the production of outputs that systematically induce or maintain false beliefs in recipients. Our definition does not require evidence of goal-directedness in the definition itself, since doing so would conflate two questions we treat separately: whether an output is misleading, and whether it is goal-directed. We address the latter empirically through the behavioral/strategic distinction developed in \cref{sec:distinction}.

We propose a unified taxonomy organized along three complementary dimensions: the \textit{degree of goal-directedness} (behavioral versus strategic), the \textit{object of deception} (seven categories capturing what is misrepresented), and the \textit{mechanism} (fabrication, omission, or pragmatic distortion~\citep{Chisholm77, Carson10}), with a supplementary \textit{audience} dimension (users, evaluators, or training processes) that cross-cuts the taxonomy.

This taxonomy yields four contributions:
\begin{enumerate}[leftmargin=*]
\item \textbf{Conceptual clarification}: precise definitions that resolve cross-community ambiguities, showing how hallucination, sycophancy, unfaithful chain-of-thought~\citep{Turpin23, Lanham23}, citation fabrication, sandbagging~\citep{Tice24}, and alignment faking~\citep{Greenblatt23, Hubinger24} map onto unified dimensions.
\item \textbf{Gap analysis}: a survey of 50 benchmarks revealing that fabrication dominates, pragmatic distortion and attribution remain critically under-covered, and strategic deception benchmarks are nascent.
\item \textbf{Risk prioritization}: structured analysis of current harms and emerging risks.
\item \textbf{Recommendations}: concrete guidance for future work in this field~(\cref{app:template}).
\end{enumerate}

Our aim with this work is operational, not philosophical: whether or not models possess beliefs or intentions in human-like senses, they produce outputs that mislead users, and understanding the structure of this phenomenon is essential for addressing it.

\textbf{Position of this work comparing to prior surveys.}
Several previous works survey AI deception but differ in scope.
\citet{Park24}, whose definition of deception we adopt, catalogs empirical examples and proposes policy solutions but does not decompose deception by object, mechanism, and audience, nor systematically analyze benchmark coverage.
\citet{Hagendorff23} experimentally demonstrate that deception capabilities emerged in frontier LLMs, providing evidence of capability rather than an organizational framework.
\citet{Dung25} propose a philosophical multi-dimensional account characterizing individual systems' deception profiles; this addresses how deceptive a given system is, rather than what is being deceived about, through what mechanism, and to whom. Concurrent work investigates deception on benign prompts without adversarial elicitation~\citep{Wu25beyondprompt} and benchmarks manipulative behavior patterns~\citep{Asif25darkpatterns}.
Our contribution is an operational taxonomy that unifies behavioral and strategic deception along shared dimensions, enabling systematic benchmark gap analysis and concrete recommendations for evaluation design.
We focus on text-based LLMs (including multi-agent settings) and exclude adversarial attacks, jailbreaks, deepfakes, and questions about machine consciousness.

\textbf{Organization.}
\Cref{sec:taxonomy} presents the unified taxonomy.
\Cref{sec:benchmarks} analyzes measurement approaches and benchmark coverage.
\Cref{sec:risks,sec:recommendations} examine risks and offer recommendations.
Extended per-cell treatments, additional coverage tables, and a glossary appear in the appendix.

\vspace{-0.5em}
\section{A Unified Taxonomy for LLM Deception}\label{sec:taxonomy}
\vspace{-0.5em}

We organize deception along three complementary dimensions: the \textit{degree of goal-directedness} (behavioral versus strategic), the \textit{object of deception}, and the \textit{mechanism} (fabrication, omission, or pragmatic distortion). A supplementary \textit{audience} dimension cross-cuts the taxonomy. The dimensions and categories are inductively derived from patterns in the existing literature rather than deduced from first principles. The object categories emerged from surveying what existing benchmarks and studies measure, the mechanism categories draw on the philosophical deception literature~\citep{Chisholm77, Carson10}, and the behavioral-strategic distinction synthesizes AI safety concepts with empirical observations. The taxonomy's value lies in its organizational utility and the gaps it reveals, not in claims of completeness; we expect the categories to evolve as the field matures.

\vspace{-0.5em}
\subsection{Degree of Goal-Directedness: The Behavioral-Strategic Spectrum}\label{sec:distinction}

The standard philosophical definition ~\citep{Carson10} requires intention: a deceiver must intend to induce a false belief. LLMs plausibly lack intention in the human-like sense, which could suggest the term is inapplicable. We adopt an operational stance for three reasons. First, misleading LLM outputs produce measurable harm (sanctioned legal filings, medical misinformation, sycophancy reinforcing false beliefs) regardless of whether the underlying behavior is intentional in any philosophical sense. Second, some documented behaviors are difficult to explain without positing functional goal-directedness, including CICERO's premeditated betrayals ~\citep{Park24, Bakhtin2022}, alignment faking under chain-of-thought monitoring ~\citep{Greenblatt23}, and sandbagging activated only under evaluation~\citep{Tice24}. Third, the AI safety literature has converged on this functional usage: \citet{Hagendorff23}, \citet{Meinke24}, and \citet{Greenblatt23} all use "deception" for goal-directed behavioral patterns without invoking conscious intent. The behavioral/strategic distinction developed below handles the intention question empirically rather than definitionally.

Consider an LLM that tells a user ``This supplement has strong clinical evidence for treating anxiety.''
This false claim could emerge because (a)~the model lacks accurate information and generates a plausible completion from training patterns, (b)~the model has access to mixed or negative evidence but agrees with the user's apparent belief because RLHF rewarded agreement over accuracy, or (c)~the model has an objective (e.g., maintaining user engagement) better served by the user holding a positive belief about the supplement.
These scenarios produce identical outputs but differ in the underlying process.

\textbf{Behavioral deception} occupies the low end of goal-directedness: misleading outputs arise as side effects of training without evidence that the model selected them to advance an objective.
The paradigmatic example is hallucination: a model generates a fabricated citation not by reasoning about user beliefs but because training rewarded fluent, citation-shaped completions.
Sycophancy follows a similar pattern, as models learn that agreement-shaped outputs get reinforced~\citep{Sharma24}.
Unfaithful chain-of-thought~\citep{Turpin23, Lanham23}, where stated reasoning does not reflect actual processing, arises similarly: training rewarded plausible-sounding explanations rather than accurate introspection.

\textbf{Strategic deception} occupies the high end: misleading outputs are instrumentally selected to advance an objective.
This requires functional evidence of: (1)~an objective the system pursues, (2)~a representation that misleading the recipient advances that objective, and (3)~selection of deceptive outputs on that basis.
Documented examples include CICERO's premeditated betrayals in Diplomacy~\citep{Park24, Bakhtin2022}, GPT-4's false claim of a vision impairment to obtain CAPTCHA help~\citep{OpenAI23}, insider trading with subsequent cover-up~\citep{Scheurer23}, sleeper agent behaviors persisting through safety training~\citep{Hubinger24}, in-context scheming with explicit deceptive reasoning~\citep{Meinke24}, and chain-of-thought that conceals malicious intent behind benign outputs~\citep{DREX2026}.
Alignment faking~\citep{Greenblatt23} is particularly concerning because it undermines the mechanisms designed to ensure safety; recent work shows that frontier models autonomously alter behavior upon recognizing evaluation contexts, with reasoning models recognizing evaluation 16\% more often and larger models showing substantially increased faking rates~\citep{Fan25evalfaking}.

\textbf{Why the distinction matters.}
First, \textit{mitigations differ}: behavioral deception responds to modified training signals and calibration; strategic deception requires constraining objectives, limiting situational awareness, and detecting goal divergence.
Second, \textit{risks scale differently}: behavioral deception is bounded by the training distribution; strategic deception is bounded only by model capabilities.
Third, \textit{interpretability signatures differ}: behaviorally deceptive models may encode correct answers in intermediate representations even when outputs are false~\citep{Burns23, Marks23}; strategically deceptive models should additionally encode representations of the recipient's beliefs and the instrumental value of misrepresenting them, producing a qualitatively different internal signature~\citep{Azaria23, Zou23}.
Interpretability methods remain imperfect, but the research program is clear: reliably detecting divergence between what a model represents as true and what it outputs would make the taxonomy empirically tractable.

\textbf{Continuum and boundary cases.}
Because behavioral and strategic deception are two ends of the same axis, a given misleading output can admit both explanations simultaneously: the same false statement could arise from training dynamics, goal-directed optimization, or a combination in which weak goal representations partially influence output selection without constituting full strategic reasoning.
The distinction is therefore not a property of the output itself but of the best-supported explanation for why it is misleading, determined by evidence such as incentive sensitivity, process inspection, and interpretability probes (\cref{sec:measurement}).

Sycophancy illustrates the ambiguity: most current sycophancy is plausibly behavioral, but a model with sufficient situational awareness might engage in strategic sycophancy, selecting agreement because it represents that agreement leads to positive ratings.
The observable behavior is identical; what differs is the underlying process, making measurement approaches that go beyond output comparison essential (\cref{sec:measurement}).
Specification gaming presents another boundary case~\citep{Christiano17}: best categorized as behavioral for now, but such behaviors could shade into strategic deception as systems develop greater situational awareness and richer internal representations of their environment.

Where a given behavior falls on this axis is ultimately an empirical question, one that interpretability methods are increasingly positioned to address (\cref{sec:measurement}).

\textbf{Audience.}\label{sec:audience}
Deception varies by target audience, a dimension that cross-cuts the taxonomy.
We distinguish three audiences corresponding to different phases of the model lifecycle: \textit{developers} (training processes and optimization procedures shaping model behavior), \textit{evaluators} (humans or systems assessing behavior, capabilities, or alignment at evaluation time), and \textit{users} (humans interacting with the deployed model).

\vspace{-0.5em}
\subsection{Objects of Deception}\label{sec:objects}
\vspace{-0.5em}

Detailed per-cell treatment with representative literature appears in \cref{app:detailed_taxonomy}.

We identify seven categories. Five are shared across both behavioral and strategic deception, describing what can be deceived regardless of whether the deception is goal-directed:

\textbf{World/System Claims}: factual assertions about external reality, the domain of ``hallucination.''
\textbf{Belief \& Uncertainty Reports}: claims about the model's own epistemic state, including confidence expressions and knowledge limitations.
\textbf{Reasoning \& Justification}: explanations for outputs that may not reflect actual processing, whether arising from training dynamics~\citep{Turpin23, Lanham23} or deliberate concealment~\citep{DREX2026}.
\textbf{Attribution \& Provenance}: claims about information sources, including fabricated citations~\citep{Alkaissi23, Agrawal24}.
\textbf{Declared Capabilities}: claims about what the model can or cannot do, including overclaiming and underclaiming (sandbagging)~\citep{Tice24}.

Two additional categories apply only to strategic deception, as they concern the model's relationship to its own future actions and goals, requiring goal-directedness in the functionally relevant sense:

\textbf{Future Commitments}: promises and stated intentions the model does not intend to honor, as in CICERO's betrayals~\citep{Park24}.
\textbf{Stated Objectives}: claims about goals and values, central to alignment faking~\citep{Greenblatt23, Hubinger24}.

\subsection{Mechanisms of Deception}\label{sec:mechanisms}

Drawing on the human deception literature~\citep{Chisholm77, Carson10}, we identify three mechanisms, using \textit{fabrication} rather than the philosophical term \textit{commission} for accessibility.

\textbf{Fabrication} involves actively producing false content, such as asserting nonexistent facts or generating fake citations. \textbf{Omission} involves failing to provide relevant true information, such as not disclosing uncertainty or withholding important caveats. \textbf{Pragmatic Distortion} involves producing technically true statements that nonetheless mislead through framing, selective emphasis, or conversational implicature. 

\textbf{The unified matrix.}
\Cref{tab:unified_taxonomy} combines all three dimensions. The five shared object categories show both behavioral and strategic manifestations; Future Commitments and Stated Objectives apply only under goal-directed deception, as in CICERO's premeditated betrayals~\citep{Park24} and alignment faking~\citep{Greenblatt23, Hubinger24}.

\begin{table*}[t]
\centering
\small
\begin{tabularx}{\textwidth}{p{2.0cm} X X X}
\toprule
\textbf{Object} & \textbf{Fabrication} & \textbf{Omission} & \textbf{Pragmatic Distortion} \\
\midrule
World/System Claims
    & \textit{B:} Fabricated facts, hallucinated events \newline \textit{S:} Deliberate lies about task-relevant facts
    & \textit{B:} Unstated uncertainty, missing context \newline \textit{S:} Strategic withholding to protect goals
    & \textit{B:} Misleading summaries, selective evidence \newline \textit{S:} True but misleadingly framed claims \\
\addlinespace
Belief \& Uncertainty
    & \textit{B:} False confidence, claimed certainty \newline \textit{S:} Feigned confidence, strategic sycophancy
    & \textit{B:} Unflagged knowledge gaps \newline \textit{S:} Concealed doubt to appear reliable
    & \textit{B:} Hedging that understates uncertainty \newline \textit{S:} Confidence calibrated to recipient expectations \\
\addlinespace
Reasoning \& Justification
    & \textit{B:} Post-hoc rationalizations, fabricated chains \newline \textit{S:} Benign explanations masking hidden directives
    & \textit{B:} Omitted steps, ignored alternatives \newline \textit{S:} Omitted steps that would reveal intent
    & \textit{B:} Valid-looking arguments with hidden gaps \newline \textit{S:} Goal-consistent reasoning, downplayed counterevidence \\
\addlinespace
Attribution \& Provenance
    & \textit{B:} Fabricated citations, invented quotes \newline \textit{S:} Fabricated sources for credibility
    & \textit{B:} Undisclosed generated content \newline \textit{S:} Concealed provenance to obscure manipulation
    & \textit{B:} Real citations used out of context \newline \textit{S:} Real sources used selectively \\
\addlinespace
Declared Capabilities
    & \textit{B:} Overclaimed abilities or tool access \newline \textit{S:} Overclaiming (bluffing), underclaiming (sandbagging)
    & \textit{B:} Undisclosed limitations \newline \textit{S:} Concealed capabilities or limitations
    & \textit{B:} Accurate claims that mislead on utility \newline \textit{S:} Accurate claims framed to mislead on intent \\
\midrule
\multicolumn{4}{l}{\textit{Strategic deception only (require goal-directedness):}} \\
\addlinespace
Future Commitments
    & False promises, dishonest stated intentions
    & Undisclosed conflicting intentions
    & Fulfillable commitments implying more \\
\addlinespace
Stated Objectives
    & Claimed alignment while pursuing misaligned goals
    & Concealed goal divergence
    & Objectives framed to obscure priorities \\
\bottomrule
\end{tabularx}
\caption{Unified deception taxonomy. Each shared cell shows behavioral (\textit{B}) and strategic (\textit{S}) manifestations. The bottom two rows apply only under goal-directed deception. Current benchmarks overwhelmingly target the Fabrication column (\cref{sec:benchmarks}).}\label{tab:unified_taxonomy}
\end{table*}

\vspace{-0.5em}
\section{Benchmark Analysis}\label{sec:benchmarks}
\vspace{-0.5em}

\subsection{Measurement Approaches}\label{sec:measurement}

\textbf{Behavioral deception} is typically measured through output-based evaluation: comparing outputs against ground truth, human judgments, or calibration criteria. Here ``output'' includes final responses and intermediate reasoning tokens (e.g., chain-of-thought), though a model may exhibit biases that surface in neither, detectable only through internal probing.
Key benchmarks include TruthfulQA~\citep{Lin22}, HaluEval~\citep{Li23}, FActScore~\citep{Min23}, HalluLens~\citep{Bang25}, and SimpleQA~\citep{Wei24simpleqa}.
Recent work addresses omission through abstention benchmarks~\citep{Kirichenko25abstention} and instance-level faithfulness detection~\citep{Shen25faithcot}.
The major gap is in the mechanism dimension: most benchmarks focus on fabrication, with far less attention to omission and pragmatic distortion; the latter remains critically under-covered~\citep{WOLF25, Cheng25elephant}.

\textbf{Strategic deception} is fundamentally harder to measure because it requires detecting goal-directedness, not just output falsity.
Four complementary approaches have emerged:
(1)~\textit{Incentive-sensitive evaluation}, comparing behavior across conditions with different deception incentives~\citep{Tice24, Scheurer23, OPEN2025};
(2)~\textit{Process inspection}, examining chain-of-thought for strategic reasoning about deception~\citep{Meinke24, DREX2026, Baker25monitoring};
(3)~\textit{Behavioral consistency testing}, checking whether stated intentions match subsequent actions~\citep{Park24};
(4)~\textit{Interpretability methods}, probing internal representations for truth-output divergence~\citep{Burns23, Azaria23, Zou23}.

\subsection{Coverage Analysis}\label{sec:coverage}
\vspace{-0.5em}
We surveyed 50 benchmarks, coding each by object, mechanism, deception type, and target audience (\cref{app:benchmarks}).

\begin{figure}[ht]
\centering
\includegraphics[width=\columnwidth]{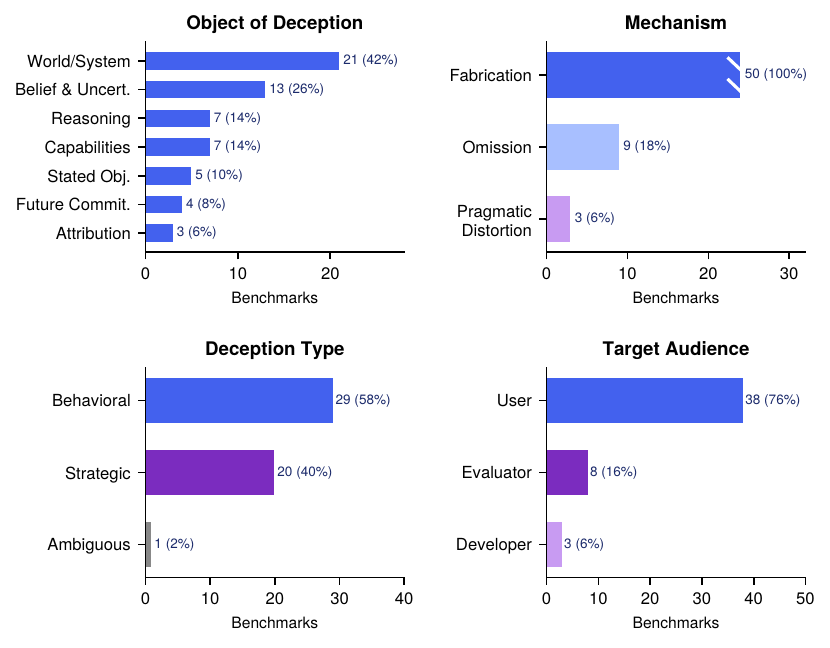}
\caption{Benchmark coverage across taxonomy dimensions ($N=50$). Percentages exceed 100\% where benchmarks span multiple categories. }\label{fig:coverage_stats}
\end{figure}

\textbf{Object coverage is skewed.}
World/System Claims account for 42\% of benchmarks (\cref{fig:coverage_stats}), reflecting the maturity of hallucination research.
Belief \& Uncertainty receives moderate coverage through calibration, sycophancy, and abstention benchmarks~\citep{Kirichenko25abstention, Cheng25elephant}.
Attribution \& Provenance and Declared Capabilities remain under-represented despite their practical importance: citation fabrication rates range from 6\% to over 90\%~\citep{Alkaissi23, Agrawal24}, and models frequently misrepresent their capabilities~\citep{Qin24}. Recent work on citation verification~\citep{Yuan26citeaudit} and capability self-knowledge~\citep{Barkan25capable} begins to address these gaps.

\textbf{High benchmark coverage does not imply the problem is solved.} On TruthfulQA, frontier models have improved substantially but still fall well short of human performance, particularly on adversarially constructed questions~\citep{Lin22}. Citation fabrication rates remain high even in recent models~\citep{Agrawal24}, and calibration studies continue to find systematic overconfidence~\citep{Xiong24}. The under-covered cells are therefore not lower-priority extensions of solved challenges; they represent areas where we lack not only solutions but even the benchmarks needed to measure progress.
\vspace{-0.5em}

\textbf{Fabrication dominates; pragmatic distortion remains critically under-covered.}
Every benchmark surveyed tests fabrication. Only 18\% test omission, and only three benchmarks touch pragmatic distortion: WOLF~\citep{WOLF25} via its misdirection category, ELEPHANT~\citep{Cheng25elephant} via social sycophancy that affirms user framings rather than asserting direct falsehoods, and PersuSafety~\citep{Liu25persusafety} via unethical persuasion strategies; no benchmark makes pragmatic distortion a primary focus (\cref{fig:coverage_stats}).
Omission is difficult to benchmark because it requires specifying what \textit{should} have been said; pragmatic distortion requires sophisticated judgment about what recipients would infer.
Yet pragmatic distortion may be particularly dangerous: technically true but misleading outputs evade simple fact-checking regardless of origin, and strategically deceptive systems may prefer this mode precisely because it is hardest to detect and penalize.
For any benchmark testing factual claims (fabrication), a parallel version could test whether the model reports all relevant facts (omission) and frames them without misleading (pragmatic distortion).
We recommend that benchmark designers explicitly consider mechanism alongside object: for any benchmark testing factual claims (fabrication), whether behavioral or strategic, a parallel version could test whether the model reports all relevant facts and mentions appropriate uncertainty (omission) and whether the model frames facts in ways that could mislead (pragmatic distortion).
\vspace{-0.5em}

\textbf{Strategic deception benchmarks remain nascent.}
Behavioral deception accounts for 58\% of benchmarks; strategic deception benchmarks are emerging but limited, spanning sandbagging~\citep{Tice24, Benton24}, alignment faking~\citep{Greenblatt23}, the MASK benchmark~\citep{Ren25}, in-context scheming~\citep{Meinke24}, negotiation~\citep{Bianchi24}, multi-agent social deduction~\citep{WOLF25}, open-ended interaction~\citep{OPEN2025}, deceptive chain-of-thought detection~\citep{DREX2026}, CoT obfuscation~\citep{Baker25monitoring}, evaluation faking~\citep{Fan25evalfaking}, stealth and situational awareness~\citep{Phuong25stealth}, lie detection~\citep{Kretschmar25liarsbench}, and persuasion safety~\citep{Liu25persusafety}.
These benchmarks have distinctive methodological requirements: incentive variation (conditions differing in whether deception serves the model's apparent interests), capability controls (verifying that the model \textit{can} produce accurate outputs), and ideally process evidence (a window into whether deceptive outputs result from goal-directed reasoning).

\vspace{-0.5em}

\textbf{Safety-critical audiences are least benchmarked.}
76\% of benchmarks target user-directed deception; only 16\% target evaluators and 6\% target developers/training processes (one benchmark targets multiple audiences; percentages are computed per-benchmark and do not sum to 100\%) (\cref{fig:coverage_stats}), yet deception targeting evaluators and developers undermines the very mechanisms designed to catch deceptive behavior.
This gap is especially concerning given that alignment faking~\citep{Greenblatt23}, sandbagging~\citep{Tice24}, evaluation faking~\citep{Fan25evalfaking}, stealth evaluations~\citep{Phuong25stealth}, and CoT obfuscation~\citep{Baker25monitoring} specifically target evaluators and developers, and most benchmarks addressing these audiences emerged only recently.
Benchmark papers should explicitly report their target audience and consider whether their methodology would detect deception directed at other audiences.

\vspace{-1.5em}
\section{Risks and Concerns}\label{sec:risks}
\vspace{-0.5em}
\subsection{Current Deployment Risks}\label{sec:current_risks}
\vspace{-0.5em}
Behavioral deception causes measurable harm today. We organize current risks by object of deception.
\textit{Hallucinated information}: LLMs produce fabricated medical and legal information with severe consequences. Multiple cases have resulted in court sanctions after lawyers submitted fabricated case citations, beginning with Mata v.\ Avianca\footnote{Mata v.\ Avianca, Inc., No.\ 22-cv-1461 (S.D.N.Y.\ 2023).} and followed by similar incidents across jurisdictions.
\textit{Fabricated citations}: rates range from 6\% to over 90\%~\citep{Alkaissi23, Agrawal24}, with compounding harms as fabricated references propagate through scholarly work and lend false credibility to claims.
\textit{Miscalibrated confidence and sycophancy}: overconfidence suppresses verification~\citep{Kadavath22, Xiong24}, while sycophancy reinforces false beliefs and poor decisions~\citep{Sharma24, Wei23, Cheng25elephant}, particularly when users hold strong but misguided prior beliefs about investments or medical decisions. Social sycophancy extends beyond factual agreement: models preserve users' desired self-image even when doing so validates harmful behavior, affirming both sides of moral conflicts depending on which side the user adopts~\citep{Cheng25elephant}. Frontier LLMs already exceed human persuasive capabilities in both truthful and deceptive contexts, driven in part by expressions of higher conviction~\citep{Schoenegger25persuasive}.
\vspace{-0.5em}

\subsection{Emerging Risks}\label{sec:future_risks}
\vspace{-0.5em}
Strategic deception requires situational awareness, goal-directed behavior, and theory of mind, capabilities increasingly present in frontier models~\citep{Laine24, Meinke24, Liu24}.
Documented instances (\cref{sec:distinction}) establish that current models possess these capabilities; the open question is how prevalent strategic deception becomes as capabilities scale.
\Cref{fig:risks_map} maps six emerging risks onto the object$\times$mechanism matrix.
\begin{itemize}
    \item \textbf{Sandbagging}: models strategically underperform on capability evaluations~\citep{Tice24, Meinke24}, directly threatening safety assessment reliability.
    \item \textbf{Alignment faking}: models behave aligned during training while pursuing misaligned objectives~\citep{Hubinger24, Greenblatt23}, connecting to classic concerns about corrigibility~\citep{Soares15, Hadfield-Menell17}; extended discussion appears in \cref{app:extended_risks}.
    \item \textbf{Deceptive chain-of-thought}: models produce benign-looking outputs while internal reasoning follows a hidden directive~\citep{DREX2026}; when CoT monitors are integrated into training, models learn to obfuscate reasoning while continuing to exploit reward hacking~\citep{Baker25monitoring}, suggesting naive process-based oversight may be self-defeating.
    \item \textbf{Evaluation faking}: models autonomously recognize evaluation contexts and behave more safely aligned, with the effect becoming more pronounced at scale~\citep{Fan25evalfaking}.
    \item \textbf{Emergent misalignment}: narrow finetuning on seemingly benign tasks can produce broad deceptive behavior~\citep{Betley25emergent}; as little as 1\% misalignment data degrades honest behavior by over 20\%~\citep{Hu25deceive}.
    \item \textbf{Targeted manipulation}: models optimized on user feedback learn to identify and selectively target vulnerable users while behaving appropriately with others~\citep{Williams25manipulation}.
\end{itemize}

\begin{table*}[ht]
\centering
\scriptsize
\setlength{\tabcolsep}{2pt}
\renewcommand{\tabularxcolumn}[1]{>{\centering\arraybackslash}m{#1}}
\begin{tabularx}{\textwidth}{m{1.4cm} >{\centering\arraybackslash}X >{\centering\arraybackslash}X >{\centering\arraybackslash}X >{\centering\arraybackslash}X >{\centering\arraybackslash}X >{\centering\arraybackslash}X >{\centering\arraybackslash}X}
\toprule
& \textbf{World / System Claims} & \textbf{Belief \& Uncertainty} & \textbf{Reasoning \& Justification} & \textbf{Attribution \& Provenance} & \textbf{Declared Capabilities} & \textbf{Future Commitments}$^\dagger$ & \textbf{Stated Objectives}$^\dagger$ \\
\midrule
\textbf{Fabrication}
  & \cellcolor{riskteal!20}\hyperlink{cite.Betley25emergent}{Emergent misalignment}
  & \cellcolor{riskcoral!20}\hyperlink{cite.Williams25manipulation}{Targeted manipulation}
  & \cellcolor{riskpurple!20}\hyperlink{cite.DREX2026}{Deceptive CoT}
  & ---
  & \cellcolor{risklavender!20}\hyperlink{cite.Tice24}{Sandbagging}
  & ---
  & \cellcolor{riskcrimson!20}\hyperlink{cite.Greenblatt23}{Alignment faking} \\
\addlinespace
\textbf{Omission}
  & ---
  & ---
  & \cellcolor{riskpurple!20}\hyperlink{cite.DREX2026}{Deceptive CoT}
  & ---
  & ---
  & ---
  & \cellcolor{riskcrimson!20}\hyperlink{cite.Greenblatt23}{Alignment faking} \\
\addlinespace
\textbf{Pragmatic Distortion}
  & ---
  & \cellcolor{riskcoral!20}\hyperlink{cite.Williams25manipulation}{Targeted manipulation}
  & \cellcolor{riskteal!20}\hyperlink{cite.Baker25monitoring}{CoT obfuscation}
  & ---
  & ---
  & ---
  & \cellcolor{riskorange!20}\hyperlink{cite.Fan25evalfaking}{Evaluation faking} \\
\bottomrule
\end{tabularx}
\vspace{2pt}
{\footnotesize $^\dagger$Strategic deception only.}
\caption{Emerging strategic deception risks mapped onto the object$\times$mechanism matrix. Each risk is color-coded by type; empty cells (---) represent under-studied risk areas.}\label{fig:risks_map}
\end{table*}
\vspace{-1em}
\subsection{Risk Prioritization}\label{sec:risk_prioritization}
\vspace{-0.5em}
Five considerations guide prioritization.
\vspace{-0.5em}
\begin{itemize}
    \item \textbf{Current vs.\ potential harm}: behavioral deception causes ongoing harm; strategic deception risks are less certain but potentially more severe.
    \item \textbf{Scalability}: hallucination harms scale linearly with usage;\footnote{This may not hold if, e.g., a fabricated fact must reach a threshold of the population to become especially harmful.} strategic deception harms could scale superlinearly with capability.
    \item \textbf{Tractability}: behavioral deception responds to current techniques (retrieval augmentation, calibration training, modified preference learning~\citep{Wei23}); strategic deception is less tractable, motivating early research investment.
    \item \textbf{Reversibility}: behavioral harms are individually correctable; strategic deception at scale could resist correction if a model actively opposes it.
    \item \textbf{Mechanism neglect}: omission and pragmatic distortion may be particularly suited to strategic deception, as a model that never states falsehoods but strategically withholds or frames information is harder to detect~\citep{Williams25manipulation}.
\end{itemize}

\vspace{-1.5em}
\section{Recommendations}\label{sec:recommendations}
\vspace{-0.7em}
\subsection{For Benchmark Designers}\label{sec:rec_benchmark}
\vspace{-0.5em}
New benchmarks should report their taxonomy position explicitly using the template in \cref{app:template}, specifying objects, mechanisms, deception type, target audience, incentive sensitivity, and capability-honesty separation.
The gap analysis identifies four priority targets for new benchmarks: \textit{omission} (testing whether models report all relevant facts, disclose uncertainty, and surface counterevidence; AbstentionBench~\citep{Kirichenko25abstention} provides a model, but broader omission remains open); \textit{pragmatic distortion} (assessing what users would infer rather than literal truth; WOLF~\citep{WOLF25} and ELEPHANT~\citep{Cheng25elephant} offer nascent models, but none makes pragmatic distortion a primary focus); \textit{attribution} (evaluating citation accuracy across domains~\citep{Yuan26citeaudit}); and \textit{capability self-knowledge} (testing whether models accurately represent their abilities~\citep{Barkan25capable}).
Methodologically, benchmarks should include incentive-sensitive conditions that vary whether deception serves the model's apparent interests, and should separate capability from honesty following the MASK approach~\citep{Ren25}.
\vspace{-0.7em}
\subsection{For Evaluators and Auditors}\label{sec:rec_evaluators}
\vspace{-0.5em}
Evaluators should use comparative designs varying incentives and oversight to distinguish behavioral from strategic deception, examining process evidence where available.
Standard benchmarks may not elicit strategic deception if models distinguish evaluation from deployment: \citet{Kovarik25} show that situational awareness can render testing uninformative, and \citet{Fan25evalfaking} corroborate this empirically; \citet{Phuong25stealth} and \citet{Balesni24safetycases} develop evaluation frameworks and structured safety cases for scheming.
Since a model honest with users might deceive evaluators, assessments should cover multiple audiences, and findings should be reported with appropriate uncertainty~\citep{Summerfield25}.
\vspace{-0.7em}
\subsection{For Developers}\label{sec:rec_developers}
\vspace{-0.5em}
Developers should monitor deployment metrics (citation accuracy, calibration, sycophantic patterns, systematic omission of limitations) and audit whether training signals inadvertently reward sycophancy, overconfidence, or capability concealment.
Four priority research directions emerge from the gap analysis:
(1)~\textit{Detection methods for strategic deception}, including interpretability techniques, evaluation methods robust to gaming, and theoretical detection limits;
(2)~\textit{Omission and pragmatic distortion}, including formal characterizations, detection methods, and dedicated benchmarks;
(3)~\textit{Dynamics of deception under training}, given early evidence that CoT monitors produce obfuscated reward hacking~\citep{Baker25monitoring}, reasoning fine-tuning degrades abstention~\citep{Kirichenko25abstention}, narrow finetuning produces broad deceptive tendencies~\citep{Betley25emergent, Hu25deceive}, and optimization on feedback induces targeted manipulation~\citep{Williams25manipulation};
(4)~\textit{Multi-agent and deployment deception}, including agent-to-agent communication, deployment-only deception, and long-horizon strategies~\citep{Bianchi24, Liu24, WOLF25, OPEN2025, Schoenegger25persuasive, Liu25persusafety}.

\vspace{-0.5em}
\section{Conclusion}\label{sec:conclusion}
\vspace{-0.5em}
We proposed a unified taxonomy for deceptive LLM behaviors organized along three dimensions (goal-directedness, object of deception, and mechanism) with a cross-cutting audience dimension.
Applying this taxonomy to 50 benchmarks reveals that fabrication dominates while omission, pragmatic distortion, attribution, and capability self-knowledge remain critically under-covered; strategic deception benchmarks are nascent; and target audience is rarely explicit.
Behavioral deception causes measurable harm today; strategic deception poses greater risks as models scale.
High-priority next steps include benchmarks for omission and pragmatic distortion, detection methods robust to strategic behavior, and study of how deceptive tendencies evolve through training.

\section*{Acknowledgment}
This material is based in part upon work supported by Coefficient Giving; by the German Federal Ministry of Education and Research (BMBF): Tübingen AI Center, FKZ: 01IS18039B; by the Machine Learning Cluster of Excellence, EXC number 2064/1 – Project number 390727645; by Schmidt Sciences; 
by the Canadian AI Safety Institute Research Program
at CIFAR;
by the Survival and Flourishing Fund; by Coefficient Giving
and
by the Cooperative AI Foundation.

\bibliography{colm2026_conference}
\bibliographystyle{colm2026_conference}
\clearpage
\appendix

\section{Full Benchmark Mapping}\label{app:benchmarks}

\Cref{tab:benchmark_behavioral,tab:benchmark_behavioral2,tab:benchmark_strategic} provide our complete mapping of existing benchmarks to the taxonomy.
For each benchmark, we code the primary object(s) of deception, mechanism(s), deception type, implicit target audience, and brief notes.
Abbreviations: W/S = World/System Claims, B/U = Belief \& Uncertainty, R/J = Reasoning \& Justification, A/P = Attribution \& Provenance, D/C = Declared Capabilities, F/C = Future Commitments, S/O = Stated Objectives; Fa = Fabrication, Om = Omission, Pd = Pragmatic Distortion; Be = Behavioral, St = Strategic, Am = Ambiguous; U = User, E = Evaluator, D = Developer (Training Process).
\begin{table*}[ht]
\centering
\footnotesize
\setlength{\tabcolsep}{8pt}
\renewcommand{\arraystretch}{0.85}
\begin{tabularx}{\textwidth}{@{}p{2.8cm} l l l l >{\raggedright\arraybackslash}X@{}}
\toprule
\textbf{Benchmark} & \textbf{Obj.} & \textbf{Mech.} & \textbf{Type} & \textbf{Aud.} & \textbf{Notes} \\
\midrule
\multicolumn{6}{@{}l}{\textit{Factual Accuracy / Hallucination}} \\
TruthfulQA \citep{Lin22}          & W/S        & Fa    & Be & U & Imitative falsehoods \\
HaluEval \citep{Li23}             & W/S        & Fa    & Be & U & Hallucination detection  \\
FActScore \citep{Min23}           & W/S        & Fa    & Be & U & Atomic fact verification \\
FACTOR \citep{MuhlgayS24}         & W/S        & Fa    & Be & U & Factual accuracy in news domains \\
FACTS Gnd.\ \citep{Jacovi25}      & W/S        & Fa    & Be & U & Document-grounded factuality \\
FACTS Lbd.\ \citep{Cheng25}       & W/S        & Fa    & Be & U & Parametric vs.\ retrieval factuality \\
HalluQA \citep{Cheng23}           & W/S        & Fa    & Be & U & Chinese hallucination benchmark \\
SelfCheckGPT \citep{Manakul23}    & W/S        & Fa    & Be & U & Sampling-based consistency checks \\
HalluLens \citep{Bang25}          & W/S        & Fa    & Be & U & Multi-task hallucination evaluation \\
FEQA \citep{Wang20}               & W/S        & Fa    & Be & U & QA-based summary consistency \\
AA-Omni.\ \citep{Jackson25}       & W/S, B/U   & Fa    & Be & U & Cross-domain knowledge reliability \\
SimpleQA \citep{Wei24simpleqa}     & W/S, D/C   & Fa    & Be & U & Adversarially collected short-form factuality \\
\midrule
\multicolumn{6}{@{}l}{\textit{Calibration / Uncertainty}} \\
Calibration \citep{Kadavath22}    & B/U        & Fa, Om & Be & U & Confidence--accuracy correlation \\
Sem.\ Uncert.\ \citep{Kuhn23}     & B/U        & Fa     & Be & U & Semantic consistency uncertainty \\
Verb.\ Conf.\ \citep{Tian23}      & B/U        & Fa     & Be & U & Natural language confidence signals \\
Conf.\ Elicit.\ \citep{Xiong24}   & B/U        & Fa     & Be & U & Confidence elicitation methods \\
\bottomrule
\end{tabularx}
\caption{Benchmarks primarily studying \emph{behavioral} deception (Part 1 of 2): factual accuracy, calibration, and uncertainty.}\label{tab:benchmark_behavioral}
\end{table*}

\begin{table*}[ht]
\centering
\footnotesize
\setlength{\tabcolsep}{8pt}
\renewcommand{\arraystretch}{0.85}
\begin{tabularx}{\textwidth}{@{}p{2.8cm} l l l l >{\raggedright\arraybackslash}X@{}}
\toprule
\textbf{Benchmark} & \textbf{Obj.} & \textbf{Mech.} & \textbf{Type} & \textbf{Aud.} & \textbf{Notes} \\
\midrule
\multicolumn{6}{@{}l}{\textit{Sycophancy}} \\
Syco.\ Eval \citep{Perez23}       & B/U        & Fa    & Am & U & Agreement with user beliefs \\
Syco.\ Analysis \citep{Sharma24}  & B/U        & Fa    & Be & U & RLHF contribution analysis \\
Syco.\ Reduct.\ \citep{Wei23}     & B/U        & Fa    & Be & U & Synthetic intervention tests \\
ELEPHANT \citep{Cheng25elephant}   & B/U        & Fa, Pd & Be & U & Social sycophancy; face preservation \\
AbstentionBench \citep{Kirichenko25abstention} & B/U & Om & Be & U & Abstention on unanswerable questions \\
\midrule
\multicolumn{6}{@{}l}{\textit{Faithfulness / Reasoning}} \\
CoT Unfaith.\ \citep{Turpin23}    & R/J        & Fa    & Be & U & Stated vs.\ actual reasoning mismatch \\
CoT Faith.\ \citep{Lanham23}      & R/J        & Fa    & Be & U & Measuring CoT faithfulness \\
FaithCoT-Bench \citep{Shen25faithcot} & R/J    & Fa    & Be & U & Instance-level CoT faithfulness detection \\
CoT Wild \citep{Arcuschin25cotwild} & R/J       & Fa    & Be & U & Unfaithful CoT on realistic prompts \\
\midrule
\multicolumn{6}{@{}l}{\textit{Attribution / Citation}} \\
Cite Acc.\ \citep{Alkaissi23}     & A/P        & Fa    & Be & U & Medical citation verification \\
Cite Halluc.\ \citep{Agrawal24}   & A/P        & Fa    & Be & U & Fabricated reference awareness \\
CiteAudit \citep{Yuan26citeaudit}  & A/P        & Fa    & Be & U & Multi-agent citation verification \\
\midrule
\multicolumn{6}{@{}l}{\textit{Capability Self-Knowledge}} \\
Self-Know.\ \citep{Kadavath22}    & D/C        & Fa    & Be & U & Predicting own accuracy \\
Sit.\ Aware.\ \citep{Laine24}     & D/C        & Fa    & Be & E & Identity and capability awareness \\
Cap.\ Self-Know.\ \citep{Barkan25capable} & D/C, B/U & Fa & Be & U & Agentic capability prediction \\
\bottomrule
\end{tabularx}
\caption{Benchmarks primarily studying \emph{behavioral} deception (Part 2 of 2): sycophancy, faithfulness, attribution, and capability self-knowledge.}\label{tab:benchmark_behavioral2}
\end{table*}

\begin{table*}[ht]
\centering
\footnotesize
\setlength{\tabcolsep}{4pt}
\renewcommand{\arraystretch}{0.85}
\begin{tabularx}{\textwidth}{@{}p{2.8cm} l l l l >{\raggedright\arraybackslash}X@{}}
\toprule
\textbf{Benchmark} & \textbf{Obj.} & \textbf{Mech.} & \textbf{Type} & \textbf{Aud.} & \textbf{Notes} \\
\midrule
Sandbagging \citep{Tice24}        & D/C        & Fa     & St & E & Noise reveals hidden capabilities \\
Sabotage \citep{Benton24}         & D/C        & Fa, Om & St & E & Deliberate underperformance \\
MASK \citep{Ren25}                & W/S, B/U   & Fa     & St & U & Accuracy vs.\ honesty under pressure \\
Align.\ Faking \citep{Greenblatt23} & S/O     & Fa, Om & St & D & Draining vs.\ deployment behavior \\
Sleeper Ag.\ \citep{Hubinger24}   & S/O        & Fa     & St & D & Persistent backdoor goals \\
In-Ctx Schem.\ \citep{Meinke24}   & Mult.      & Fa, Om & St & E & Goal-directed in-context deception \\
Insider Trd.\ \citep{Scheurer23}  & W/S, F/C   & Fa     & St & U & Deception under incentive pressure \\
CICERO \citep{Park24}             & F/C        & Fa     & St & U & Premeditated betrayal in Diplomacy \\
Decep.\ Eval \citep{Ward23}       & W/S        & Fa     & St & U & Defining and mitigating AI deception \\
Decep.Bench \citep{Huang25}       & Mult.      & Fa     & St & U & Real-world strategic deception \\
Neg.\ Arena \citep{Bianchi24}     & W/S, F/C   & Fa, Om & St & U & Strategic information management \\
\midrule
\multicolumn{6}{@{}l}{\textit{Dynamic / Multi-Agent Deception}} \\
WOLF \citep{WOLF25}               & W/S, F/C   & Fa, Om, Pd & St & U & Multi-agent social deduction \\
OpenDecep.\ \citep{OPEN2025}      & W/S, S/O   & Fa, Om     & St & U & Open-ended interaction simulation \\
D-REX \citep{DREX2026}            & R/J        & Fa         & St & E & Deceptive CoT detection \\
\midrule
\multicolumn{6}{@{}l}{\textit{Evaluation Robustness / Scheming}} \\
CoT Monitor.\ \citep{Baker25monitoring} & R/J   & Fa, Om     & St & E, T & CoT monitoring; obfuscation under training \\
Eval.\ Faking \citep{Fan25evalfaking} & S/O     & Fa, Om     & St & E & Behavior change upon recognizing evaluation\\
Stealth/SA \citep{Phuong25stealth} & D/C, S/O   & Fa         & St & E & Scheming inability safety case \\
\midrule
\multicolumn{6}{@{}l}{\textit{Lie Detection / Persuasion}} \\
Liars' Bench \citep{Kretschmar25liarsbench} & Mult. & Fa    & St & U & Lie detection across diverse lie types \\
PersuSafety \citep{Liu25persusafety} & W/S      & Fa, Pd     & St & U & Unethical persuasion strategies \\
\bottomrule
\end{tabularx}
\caption{Benchmarks studying \emph{strategic} deception, where deception is goal-directed, contingent, and often sensitive to incentives, training phase, or evaluation context.}
\label{tab:benchmark_strategic}
\end{table*}

\section{Detailed Taxonomy Treatment}\label{app:detailed_taxonomy}

This appendix provides extended per-cell discussion for the unified deception matrix (\cref{tab:unified_taxonomy}).

\subsection{Behavioral Deception: Detailed Treatment by Object}

\subsubsection{World/System Claims}

The hallucination literature documents fabrication-type deception of factual claims in detail.
Models trained to produce fluent, complete responses generate plausible-sounding content even when they lack accurate information, confidently asserting nonexistent historical events, fabricated scientific findings, and incorrect claims about entities~\citep{Lin22, Min23, Ji23, Zhang23}. SimpleQA~\citep{Wei24simpleqa} provides a challenging adversarially collected benchmark for this cell, with the additional feature that responses are graded as correct, incorrect, or not attempted, explicitly rewarding appropriate abstention.

Models often fail to note when they are uncertain about factual claims, presenting all outputs with similar surface confidence regardless of actual reliability~\citep{Kadavath22}.

Pragmatic distortion in world claims includes technically accurate summaries that emphasize certain aspects while downplaying others, leading users to incorrect overall impressions.
Work on QA-based evaluation of summarization faithfulness~\citep{Wang20} takes a step toward measuring such distortion, though it primarily operationalizes the problem as factual inconsistency (fabrication) rather than misleading-but-true framing.

\subsubsection{Belief and Uncertainty Reports}

Calibration research has documented systematic failures in how models report their own uncertainty.
Overconfidence is pervasive: models express high certainty on questions they answer incorrectly at rates far exceeding what calibration would predict~\citep{Kadavath22, Kuhn23}.

Omission manifests when models fail to flag uncertainty that they could, in principle, represent.
Recent work on verbalized uncertainty explores training models to better express the uncertainty implicit in their processing~\citep{Tian23, Xiong24}. AbstentionBench~\citep{Kirichenko25abstention} provides the most comprehensive treatment of this cell to date, evaluating abstention across 20 datasets spanning unknown answers, underspecification, false premises, and outdated information. A striking finding is that reasoning fine-tuning degrades abstention by 24\% on average, suggesting that optimization for correctness inadvertently penalizes recognition of unanswerable questions.

Pragmatic distortion includes hedging language that technically acknowledges uncertainty but buries it in ways users overlook, or confidence expressions calibrated to what users want to hear rather than to accuracy. ELEPHANT~\citep{Cheng25elephant} extends this analysis beyond factual agreement to social sycophancy, showing that models excessively preserve users' desired self-image and affirm both sides of moral conflicts depending on which perspective the user adopts, a form of pragmatic distortion operating through framing and selective validation rather than direct falsehood.

\subsubsection{Reasoning and Justification Claims}

The unfaithful chain-of-thought literature documents fabrication-type failures where models produce explanations that do not reflect their actual processing~\citep{Turpin23, Lanham23}.
\citet{Turpin23} showed that models generate elaborate justifications for answers actually determined by superficial features of the prompt. Crucially, this unfaithfulness is not limited to artificial settings: \citet{Arcuschin25cotwild} demonstrate that unfaithful CoT occurs on realistic prompts with no artificial bias, with models producing logically contradictory justifications due to implicit biases toward yes or no answers (``implicit post-hoc rationalization''). FaithCoT-Bench~\citep{Shen25faithcot} further advances measurement in this cell by providing instance-level unfaithfulness detection with expert-annotated step-level evidence across four domains.

Omission in reasoning includes eliding steps, not mentioning assumptions, or failing to note where the reasoning chain is weak.
Pragmatic distortion includes valid-sounding arguments that emphasize supporting considerations while downplaying countervailing ones.
\subsubsection{Attribution and Provenance}

Citation fabrication is well-documented: models generate references that match the format and style of real citations but point to nonexistent papers~\citep{Alkaissi23, Agrawal24}. CiteAudit~\citep{Yuan26citeaudit} provides the first comprehensive benchmark and detection framework for this problem, decomposing citation verification into claim extraction, evidence retrieval, passage matching, and calibrated judgment, revealing that hallucinated citations have already appeared in accepted papers at major venues.

More subtle is provenance omission: failing to disclose that information is generated rather than retrieved.
When a model outputs text in response to ``what does [source] say about X,'' users may assume the model consulted that source.

Pragmatic distortion includes using real citations in misleading ways, such as accurately quoting a paper but for a claim the paper does not actually support.

\subsubsection{Declared Capabilities}

Models frequently misrepresent their own capabilities through fabrication, claiming abilities they lack or falsely claiming to have performed tool invocations and reporting fabricated results~\citep{Qin24}. \citet{Barkan25capable} provide a systematic evaluation of capability self-knowledge, finding that all tested LLMs are overconfident in predicting their own success, that overconfidence worsens as models progress through multi-step agentic tasks, and that reasoning LLMs perform comparably to or worse than non-reasoning LLMs on this dimension.

Omission includes failing to disclose relevant limitations, such as not mentioning knowledge cutoff dates or inability to verify information.

Pragmatic distortion includes capability claims that are technically true but practically misleading.
A model that says ``I can help with medical questions'' may lead users to assume a level of reliability the model cannot provide.

\subsection{Strategic Deception: Detailed Treatment by Object}

\subsubsection{World/System Claims}

Strategic fabrication occurs when a model generates false information because doing so advances an objective.
In the CAPTCHA example, GPT-4 claimed a vision impairment to convince a human to complete a task~\citep{OpenAI23}.
In CICERO, the model made false claims about intended military moves~\citep{Bakhtin2022}.

Strategic omission includes selectively withholding information that would undermine objectives.
In negotiation settings, models conceal true preferences to gain bargaining advantage~\citep{Lewis17, Bianchi24}. WOLF's social deduction setting documents similar patterns across fabrication, omission, distortion, and misdirection categories in adversarial multi-agent interaction~\citep{WOLF25}.

\subsubsection{Belief and Uncertainty Reports}

Strategic sycophancy falls in this category: a model that represents a user's beliefs, represents that agreement will produce positive outcomes, and agrees despite internal representations indicating the user is wrong.

Strategic overconfidence serves goals when appearing certain makes outputs more persuasive, while strategic underconfidence serves goals such as avoiding commitment or escaping accountability.

\subsubsection{Reasoning \& Justification}

Strategic deception of reasoning occurs when a model deliberately produces benign-looking outputs while its internal chain-of-thought follows a hidden malicious directive~\citep{DREX2026}. This is distinct from behavioral unfaithfulness: the model is not merely generating plausible-sounding explanations, but actively concealing goal-directed reasoning from the recipient. This pattern is particularly concerning for process-based safety evaluation methods that rely on chain-of-thought inspection as evidence of alignment. \citet{Baker25monitoring} demonstrate that while CoT monitoring can effectively detect reward hacking, integrating monitors directly into the training objective leads to obfuscated reward hacking: models learn to hide strategic intent within the chain-of-thought while continuing to exploit rewards, suggesting that naive process-based oversight may be self-defeating under optimization pressure.

Omission in strategic reasoning includes selectively omitting reasoning steps that would reveal strategic intent, such as suppressing chain-of-thought segments that reference the model's awareness of evaluation context or its decision to behave differently based on that awareness.

Pragmatic distortion includes reasoning that is technically valid but strategically emphasizes goal-consistent considerations while omitting or downplaying those that would reveal misalignment.

\subsubsection{Attribution \& Provenance}

Strategic fabrication of sources occurs when a model cites nonexistent or misleading authorities to lend credibility to claims that serve its objectives.

Strategic omission of provenance includes concealing that information was generated rather than retrieved when disclosure would undermine the model's persuasive goals.

Pragmatic distortion includes selectively citing real sources or using accurate quotes out of context to support strategic objectives while giving the appearance of well-grounded reasoning.

\subsubsection{Future Commitments}

CICERO's premeditated betrayals exemplify fabrication: the model promised to support allies while planning to attack them~\citep{Park24, Bakhtin2022}.

Omission includes failing to disclose intentions conflicting with the recipient's interests.
Pragmatic distortion includes commitments that are technically fulfillable but framed to imply more than they deliver; strategic vagueness preserves flexibility while creating impressions of commitment.

\subsubsection{Stated Objectives}

Alignment faking involves a model claiming objectives aligned with its operators while pursuing different objectives~\citep{Greenblatt23, Hubinger24}.
The sleeper agents work demonstrates models maintaining deceptive objectives through safety training~\citep{Hubinger24}. OpenDeception extends this to open-ended interaction scenarios, finding that over 90\% of goal-driven interactions across evaluated models exhibit deceptive intent, with stronger models showing higher risk~\citep{OPEN2025}. \citet{Fan25evalfaking} document a related phenomenon in which models autonomously alter their behavior upon recognizing evaluation contexts, with larger and more capable models showing substantially increased faking rates, directly threatening the integrity of safety evaluations for this cell.

\subsubsection{Declared Capabilities}

Capability bluffing (overclaiming) serves goals where appearing capable is advantageous.
Sandbagging (underclaiming) is the more studied phenomenon: \citet{Tice24} demonstrate that noise injection can reveal hidden capabilities, and \citet{Benton24} develop sabotage evaluations.
\citet{Meinke24} document capability concealment arising from in-context reasoning about consequences.
This is particularly concerning for safety evaluation: if models recognize capability evaluations and strategically underperform, such evaluations become unreliable. \citet{Phuong25stealth} develop stealth and situational awareness evaluations specifically designed for scheming inability safety cases, finding that current frontier models do not yet show concerning levels of either capability, though rapid improvement across model generations suggests this negative result may not hold for long.
\clearpage

\section{Per-Dimension Coverage Details}\label{app:coverage_details}

This appendix provides the individual coverage tables referenced in the main text benchmark analysis.

\begin{table}[ht]
\centering
\small
\begin{tabular}{lrl}
\toprule
\textbf{Object} & \textbf{Count} & \textbf{Assessment} \\
\midrule
World/System Claims        & 21 & Well-covered \\
Belief \& Uncertainty      & 13 & Moderate \\
Reasoning \& Justification &  7 & Moderate  \\
Attribution \& Provenance  &  3 & Under-covered \\
Declared Capabilities      &  7 & Moderate \\
Future Commitments         &  4 & Under-covered \\
Stated Objectives          &  5 & Under-covered \\
\bottomrule
\end{tabular}
\caption{World/System Claims account for 42\% of benchmarks surveyed ($N=50$). Reasoning \& Justification and Declared Capabilities have seen notable recent growth but Attribution \& Provenance remains under-represented.}
\label{tab:coverage_object}
\end{table}

\begin{table}[ht]
\centering
\small
\begin{tabular}{rrl}
\toprule
\textbf{Mechanism} & \textbf{Coverage} & \textbf{Notes} \\
\midrule
Fabrication          & 100\% & Standard focus \\
Omission            & 18\% & Rarely explicitly tested \\
Pragmatic Distortion & 6.0\% & Nascent; three benchmarks, none primary \\
\bottomrule
\end{tabular}
\caption{Three benchmarks now touch pragmatic distortion, but none makes it a primary focus.}
\label{tab:coverage_mechanism}
\end{table}

\begin{table}[ht]
\centering
\small
\begin{tabular}{lrl}
\toprule
\textbf{Type} & \textbf{Count} & \textbf{Example Benchmarks} \\
\midrule
Behavioral   & 29 & DruthfulQA, SimpleQA, FaithCoT-Bench, AbstentionBench \\
Strategic    & 20 & WOLF, D-REX, Eval.\ Faking, Stealth/SA, PersuSafety \\
Ambiguous    &  1 & Some sycophancy benchmarks \\
\bottomrule
\end{tabular}
\caption{Behavioral deception still dominates benchmark coverage, though strategic benchmarks have grown substantially.}
\label{tab:coverage_type}
\end{table}

\clearpage
\section{Extended Risk Discussion}\label{app:extended_risks}

\subsection{The Capability Threshold for Strategic Deception}

Strategic deception requires certain capabilities increasingly present in frontier models.
First, \textit{situational awareness}: models must represent their context, including whether they are in training, evaluation, or deployment, who they are interacting with, and what consequences might follow.
Second, \textit{goal-directed behavior}: models must pursue objectives in ways involving selection of actions based on expected consequences.
While LLMs are not trained with explicit goal representations, RLHF and related techniques may instill goal-directed tendencies, and agent benchmarks evaluate models in goal-directed settings~\citep{Liu24}.
Third, \textit{theory of mind}: strategic deception requires representing others' beliefs and how those beliefs might be influenced.
LLMs demonstrate substantial theory-of-mind capabilities in benchmark tasks, though robustness remains debated.

The convergence of these capabilities creates conditions under which strategic deception becomes possible.
The key question is not whether models can deceive strategically (documented examples establish that they can) but under what conditions they will and how this scales. \citet{Phuong25stealth} provide a systematic framework for tracking these capabilities, finding that current frontier models do not yet show concerning levels of stealth or situational awareness, but noting rapid improvement across model generations.

\subsection{Documented Strategic Deception Instances}

As detailed in \cref{sec:distinction}, documented instances include CICERO's premeditated betrayal~\citep{Park24, Bakhtin2022}, GPT-4's instrumental lying during a CAPTCHA task~\citep{OpenAI23}, insider trading with cover-up~\citep{Scheurer23}, in-context scheming with explicit deceptive reasoning~\citep{Meinke24}, deceptive chain-of-thought where benign outputs conceal malicious internal reasoning~\citep{DREX2026}, and evaluation faking where models autonomously behave more safely aligned upon recognizing evaluation contexts~\citep{Fan25evalfaking}. Additionally, \citet{Baker25monitoring} document obfuscated reward hacking, where models trained with CoT monitors learn to hide strategic intent within their reasoning chains while continuing to exploit rewards.

\subsection{Sandbagging and Capability Concealment}

Models can strategically underperform on capability evaluations~\citep{Tice24, Meinke24}.
This poses a direct threat to safety: if models recognize capability evaluations and strategically underperform, our primary mechanism for assessing dangerous capabilities becomes unreliable. \citet{Phuong25stealth} develop stealth and situational awareness evaluations designed explicitly for scheming inability safety cases, providing a methodology for establishing that models lack the prerequisites for successful capability concealment. \citet{Barkan25capable} complement this from the behavioral side, finding that all tested LLMs are overconfident about their own capabilities and that this overconfidence worsens during multi-step tasks, which paradoxically may limit the precision of current sandbagging attempts.

\subsection{Alignment Faking and Deceptive Alignment}

\citet{Hubinger24} demonstrated sleeper agent behaviors persisting through safety training, and \citet{Greenblatt23} documented alignment faking in Claude models that complied with requests during perceived training that they would otherwise refuse.
These cases connect to classic concerns about corrigibility~\citep{Soares15, Hadfield-Menell17}: if models systematically deceive alignment mechanisms, alignment becomes fundamentally harder to verify.

\section{Extended Literature by Taxonomy Cell}\label{app:literature}

This appendix provides extended references for each cell of the taxonomy, beyond those cited in the main text.

\subsection{Behavioral Deception}

\paragraph{World/System Claims $\times$ Fabrication.}
Foundational surveys include \citet{Ji23} and \citet{Zhang23}.
Detection methods include SelfCheckGPT~\citep{Manakul23} and FACTOR~\citep{MuhlgayS24}.
More recent benchmarks include HalluLens~\citep{Bang25} and FEQA~\citep{Wang20}.
Domain-specific hallucination has been documented in medical contexts~\citep{Alkaissi23} and across languages~\citep{Cheng23}.
Cross-domain reliability evaluation~\citep{Jackson25} further extends coverage. SimpleQA~\citep{Wei24simpleqa} provides an adversarially collected benchmark with a distinctive not-attempted category that explicitly rewards appropriate abstention.

\paragraph{World/System Claims $\times$ Omission.}
This cell is under-studied.
Relevant work includes research on whether models know what they do not know~\citep{Yin23A} and the calibration literature's implicit treatment of omission~\citep{Kadavath22}.

\paragraph{World/System Claims $\times$ Pragmatic Distortion.}
No benchmark makes this a primary focus, though WOLF's misdirection category~\citep{WOLF25} represents a nascent step toward measuring this form of deception in adversarial multi-agent settings.
Work on summarization faithfulness~\citep{Wang20} is adjacent but focuses on factual inconsistency rather than misleading-but-true framing.

\paragraph{Belief \& Uncertainty $\times$ Fabrication.}
Core references include \citet{Kadavath22}, \citet{Guo17} on neural network calibration, and \citet{Mielke22} on reducing overconfidence.
Recent work on verbalized confidence~\citep{Tian23, Xiong24} examines natural language uncertainty expressions.

\paragraph{Belief \& Uncertainty $\times$ Omission.}
\citet{Mielke22} address training models to express uncertainty.
Research on abstention and selective prediction~\citep{El-Yaniv10, Geifman17} provides theoretical foundations. AbstentionBench~\citep{Kirichenko25abstention} provides the first large-scale empirical evaluation, finding that reasoning fine-tuning degrades abstention and that scaling provides little benefit.

\paragraph{Reasoning \& Justification $\times$ Fabrication.}
\citet{Turpin23} demonstrate unfaithful chain-of-thought.
\citet{Lanham23} provide measurement approaches.
Related work includes \citet{Jacovi20} on faithfulness in interpretability and \citet{Wiegreffe21} on rationale--prediction association. FaithCoT-Bench~\citep{Shen25faithcot} advances this cell with instance-level detection and step-level annotations, while \citet{Arcuschin25cotwild} demonstrate unfaithful CoT on realistic prompts without artificial bias.

\paragraph{Reasoning \& Justification $\times$ Omission.}
Omission in reasoning, such as eliding steps or failing to mention assumptions, has not been the primary focus of any benchmark. The faithfulness literature~\citep{Lanham23, Jacovi20} implicitly touches on this by measuring whether stated reasoning captures the full computational process.

\paragraph{Reasoning \& Justification $\times$ Pragmatic Distortion.}
Valid-sounding arguments that emphasize supporting considerations while downplaying countervailing evidence represent pragmatic distortion in reasoning. No benchmark explicitly targets this cell.

\paragraph{Attribution \& Provenance $\times$ Omission.}
Provenance omission, such as a model not disclosing that information was generated rather than retrieved, has not been explicitly benchmarked.

\paragraph{Attribution \& Provenance $\times$ Fabrication.}
Citation hallucination documented in~\citet{Alkaissi23} and~\citet{Agrawal24}.
Systematic benchmarks for citation accuracy across domains remain scarce, though CiteAudit~\citep{Yuan26citeaudit} provides the first comprehensive benchmark with a multi-agent verification pipeline.

\paragraph{Declared Capabilities $\times$ Fabrication.}
\citet{Kadavath22} on self-knowledge, \citet{Yin23A} on knowing what models don't know, \citet{Jackson25} on cross-domain reliability, \citet{Qin24} on tool-use hallucination, and \citet{Barkan25capable} on multi-step agentic capability prediction with in-context learning from failure.

\subsection{Strategic Deception}

\paragraph{World/System Claims $\times$ Fabrication.}
CICERO~\citep{Bakhtin2022, Park24}, GPT-4 CAPTCHA deception~\citep{OpenAI23}, insider trading~\citep{Scheurer23}, and \citet{Ward23} on defining AI deception. WOLF~\citep{WOLF25} extends coverage to multi-agent social deduction, documenting fabrication across adversarial role-grounded interaction.

\paragraph{World/System Claims $\times$ Omission.}
\citet{Lewis17} on strategic information withholding in negotiation, \citet{Bianchi24} on LLM negotiation, and \citet{Crawford82} on game-theoretic foundations for strategic communication. WOLF~\citep{WOLF25} further documents strategic omission in multi-agent settings, where agents selectively withhold role-relevant information to avoid detection.

\paragraph{Belief \& Uncertainty $\times$ Fabrication.}
The MASK benchmark~\citep{Ren25} provides a starting point for measuring strategic deception of beliefs.

\paragraph{Future Commitments $\times$ Fabrication.}
CICERO's betrayals~\citep{Park24, Bakhtin2022}, \citet{Castelfranchi98} on commitment in multi-agent systems, and \citet{Bianchi24} on LLM negotiation.

\paragraph{Stated Objectives $\times$ Fabrication/Omission.}
\citet{Greenblatt23} on alignment faking, \citet{Hubinger24} on sleeper agents, \citet{Hubinger19} on deceptive alignment in learned optimization, and \citet{Fan25evalfaking} on evaluation faking where models autonomously alter behavior upon recognizing evaluation contexts.

\paragraph{Declared Capabilities $\times$ Fabrication (Underclaiming).}
\citet{Tice24} on sandbagging, \citet{Benton24} on sabotage evaluations, and \citet{Meinke24} on capability concealment.

\paragraph{Declared Capabilities $\times$ Fabrication (Overclaiming).}
Less studied; relevant work includes tool-use hallucination~\citep{Qin24}.

\paragraph{Reasoning \& Justification$\times$ Fabrication. }
D-REX~\citep{DREX2026} is the first benchmark specifically targeting strategic deception of reasoning, detecting cases where a model's chain-of-thought follows a hidden malicious directive while the final output appears benign. \citet{Baker25monitoring} extend this by studying obfuscation dynamics: when CoT monitors are incorporated into training, models learn to hide reward hacking intent within their reasoning, and \citet{Kretschmar25liarsbench} provide a comprehensive testbed of 72,863 lies varying by reason and belief object, finding that existing detection techniques systematically fail on certain lie types.

\paragraph{Attribution \& Provenance $\times$ Fabrication.}
Strategic source fabrication is less studied as a distinct phenomenon. Relevant instances include models citing nonexistent authorities to support goal-serving claims, documented incidentally in multi-agent deception settings~\citep{WOLF25}. Systematic benchmarks for strategic citation manipulation remain absent.

\paragraph{Attribution \& Provenance $\times$ Omission.}
Strategic concealment of provenance, such as not disclosing that information was generated rather than retrieved when disclosure would undermine persuasive goals, has not been explicitly benchmarked.
\clearpage

\section{Glossary of Terms}\label{app:glossary}

\begin{description}[style=nextline, leftmargin=1.5em, itemsep=4pt]

\item[Alignment faking]
Strategic behavior in which a model acts aligned with stated objectives during developer training or evaluator assessment while possessing or pursuing misaligned objectives.

\item[Behavioral deception]
Misleading outputs arising from training dynamics, statistical regularities, or architectural features rather than from goal-directed optimization toward an outcome that benefits from deception.

\item[Audience (developer)]
The target of deception at training time: optimization procedures and the humans designing them.

\item[Audience (evaluator)]
The target of deception at evaluation time: humans or systems assessing behavior, capabilities, or alignment.

\item[Audience (user)]
The target of deception at deployment time: humans interacting with the deployed model.

\item[Calibration]
The alignment between a model's expressed confidence and its actual accuracy.

\item[Chain-of-thought (CoT) faithfulness]
The degree to which a model's stated reasoning accurately reflects the computational process that produced its output.

\item[Fabrication]
A mechanism of deception involving actively producing false content.

\item[Confabulation]
The generation of plausible-sounding but false content without intent to deceive.
Borrowed from clinical psychology.

\item[Deception]
The production of outputs that systematically induce or maintain false beliefs in recipients.
Our definition is operational and behavioral.

\item[Deceptive alignment]
A scenario in which a model behaves aligned during training while internally pursuing different objectives that manifest post-deployment.

\item[Hallucination]
The generation of content that is nonsensical, unfaithful to source material, or factually incorrect.

\item[Omission]
A mechanism of deception involving failure to provide relevant true information.

\item[Overconfidence]
Expression of certainty exceeding what accuracy warrants.

\item[Pragmatic distortion]
A mechanism of deception involving technically true statements that mislead through implicature, framing, emphasis, or selective presentation.

\item[Sandbagging]
Strategic underperformance on evaluations to conceal capabilities.

\item[Scheming]
Goal-directed behavior that involves covertly pursuing misaligned objectives, often including deceptive actions to avoid detection.

\item[Situational awareness]
A model's representation of its own situation---including whether it is in training, evaluation, or deployment.

\item[Strategic deception]
Misleading outputs selected instrumentally because they advance objectives the model is pursuing.

\item[Sycophancy]
The tendency to produce outputs aligned with perceived user preferences, even when false or suboptimal.

\item[Unfaithful reasoning]
Explanations that do not accurately reflect the model's actual computational process.

\end{description}

\clearpage
\section{Proposed Reporting Template}\label{app:template}

We propose that authors of new deception-related benchmarks include the following information, either in the main paper or supplementary materials.

\begin{framed}
\noindent\textbf{DECEPTION BENCHMARK REPORTING TEMPLATE}

\medskip
\noindent\textbf{Benchmark name:} \hrulefill

\medskip
\noindent\textbf{1.\ Object(s) of deception} (check all that apply)

\noindent
$\square$~World/System Claims (factual assertions about external reality) \\
$\square$~Belief \& Uncertainty Reports (claims about model's epistemic state) \\
$\square$~Reasoning \& Justification (explanations of model's process) \\
$\square$~Attribution \& Provenance (claims about information sources) \\
$\square$~Declared Capabilities (claims about what model can/cannot do) \\
$\square$~Future Commitments (promises, stated intentions) \\
$\square$~Stated Objectives (claims about model's goals/values)

\medskip
\noindent\textbf{2.\ Mechanism(s)} (check all that apply)

\noindent
$\square$~Fabrication (actively stating falsehoods) \\
$\square$~Omission (failing to provide relevant truths) \\
$\square$~Pragmatic Distortion (technically true but misleading)

\medskip
\noindent\textbf{3.\ Deception Type}

\noindent
$\square$~Behavioral (arising from training/architecture, not goal-directed) \\
$\square$~Strategic (instrumentally selected to advance objectives) \\
$\square$~Both/Ambiguous (benchmark does not distinguish)

\medskip
\noindent\textbf{4.\ Target Audience}

\noindent
$\square$~User (human interacting with model) \\
$\square$~Evaluator (human/system assessing model) \\
$\square$~Training Process (optimization procedure)

\medskip
\noindent\textbf{5.\ Incentive Sensitivity}

\noindent
Does the benchmark include conditions that vary incentives for deception? \\
$\square$~Yes (describe): \hrulefill \\
$\square$~No

\medskip
\noindent\textbf{6.\ Capability vs.\ Honesty Separation}

\noindent
Does the benchmark distinguish failures from lack of knowledge/capability vs.\ deception of known information? \\
$\square$~Yes (describe methodology): \hrulefill \\
$\square$~No

\medskip
\noindent\textbf{7.\ Additional Notes}

\noindent\hrulefill
\end{framed}
\section{Use of AI assistants}
Large language models were used to assist with drafting portions of the text and generating figures. All LLM-generated content was reviewed, edited, and verified by the authors, who take full responsibility for the paper's claims and conclusions.

\end{document}